\documentclass[iop]{emulateapj}
\usepackage[rightcaption]{sidecap}
\usepackage{amsmath}
\usepackage{textcase}
\usepackage{soul}

\shorttitle{\texttt{P\MakeLowercase{and}E\MakeLowercase{xo}}}
\shortauthors{Batalha \emph{et al.}}

\begin{document}

\title{\texttt{P\MakeLowercase{and}E\MakeLowercase{xo}}: A Community Tool for Transiting \\Exoplanet Science
with JWST \& HST}

\author{Natasha E. Batalha\altaffilmark{1,2}}
\affil{Astronomy \& Astrophysics Department, Pennsylvania State University, University Park, PA 16802}
\email{neb149@psu.edu}

\author{Avi Mandell}
\affil{Solar System Exploration Division, NASA Goddard Space Flight Center, Greenbelt, MD 20771}

\author{Klaus Pontoppidan, Kevin B. Stevenson, Nikole K. Lewis, Jason Kalirai, Nick Earl}
\affil{Space Telescope Science Institute, Baltimore, MD 21218}
 
\author{Thomas Greene}
\affil{NASA Ames Research Center, Space Science and Astrobiology Division, M.S. 245-6, Moffett Field, CA 94035}

\author{Lo\"{\i}c Albert} 
\affil{Institut de recherche sur les exoplan\`etes (iREx), Universit\'e de Montr\'eal, Montr\'eal, Qu\'ebec, Canada.}

\and

\author{Louise D. Nielsen} 
\affil{Observatoire de Gen\`eve, Universit\`e de Gen\`eve, 51 Ch. des Maillettes, 1290, Sauverny, Switzerland}

\altaffiltext{1}{Center for Exoplanets and Habitable Worlds, University Park, PA}
\altaffiltext{2}{Vising student collaborator, Solar System Exploration Division, NASA Goddard Space Flight Center, Greenbelt, MD 20771}

\begin{abstract}
As we approach the James Webb Space Telescope (JWST) era, 
several studies have emerged that aim to: 1) characterize how the instruments will perform
and 2) determine what atmospheric spectral features could theoretically be detected using transmission and emission spectroscopy. To some degree, all these studies have relied on modeling of JWST's theoretical instrument noise. With under
two years left until launch, it is imperative that the exoplanet
community begins to digest and integrate these studies into their observing
plans, as well as think about how to leverage the Hubble Space Telescope (HST) to optimize JWST observations. In order to encourage this and to allow all members of the
community access to JWST \& HST noise simulations, we present here an open-source Python package and online interface
for creating observation simulations of all observatory-supported time-series
spectroscopy modes. This noise simulator, called \texttt{PandExo}, relies on some 
aspects of Space Telescope Science Institute's Exposure Time Calculator, \texttt{Pandeia}. We 
describe \texttt{PandExo} and the formalism for computing noise sources for JWST. Then, we benchmark \texttt{PandExo}'s performance against each instrument 
team's independently written noise simulator for JWST, and previous observations for HST. We find that \texttt{PandExo} is within 10\% agreement for HST/WFC3 and for all JWST instruments.
\end{abstract}


\section{Introduction}

JWST is equipped with a 6.5-meter primary mirror
and four visible to mid-IR 
instruments  
(NIRCam, NIRISS, NIRSpec, MIRI) that span 0.6-28
$\micron$ with low and medium resolution modes, which has the potential for ground-breaking
exoplanet science. This led to several studies focused on characterizing the observatory's expected performance
and estimating the planetary properties that could be constrained. 

\citet{gre07} were among the first to baseline the performance of JWST's primary imaging instrument, NIRCam, with regards to exoplanet science. They found that with
1000 seconds of integration time, R=500 spectra of Jupiter-sized exoplanets in primary
transit and secondary eclipse will be attainable with signal-to-noise ratios (SNRs) ranging
from $\sim$5 for faint (M=10 mag) G2V stars and up to $\sim$90 for bright (M=5 mag)
G2V stars.  \citet{dem09} created a sensitivity model for NIRSpec and MIRI, and predicted
that JWST will be able to measure temperature and absorption of CO$_2$ and
H$_2$O in 1-4 habitable Earth-like planets discovered by the \emph{Transiting
Exoplanet Survey Satellite} (TESS). 

Since then, many have sought to baseline the performance of JWST using independent sensitivity models \citep{kal09,bei14,bat15, cow15,bar15, bar16, gre16, mor16,mol16,how17,bat17}. For example, \citet{bat15} reported that primary transit spectroscopy with
NIRSpec of 1-10 M$_\earth$, 400-1000 K planets orbiting M dwarfs would
result in high SNR spectra if the planets were within $\sim$50 pc and if 25
transits were co-added. These results were based off noise simulations which
included spacecraft jitter, drift, flat field errors and background noise. In
reality, exoplanet observations with JWST could suffer from other systematics
as well. \citet{bar15} explored these effects by including time varying
astrophysical and instrumental systematics in their observational simulations. \citet{gre16} used
a retrieval algorithm with an independent noise simulator for NIRISS,
NIRCam, and MIRI in order to determine what atmospheric properties could be
retrieved from a hot Jupiter, warm Neptune, warm sub-Neptune and cool
super-Earth. All of these were pivotal to our knowledge and understanding of the functionality of JWST observing modes and all of these relied on simulating noise sources. 

With just two years left until launch, it is imperative that the exoplanet
community begins to digest and integrate these studies into their observing
plans and strategies. In order to encourage this and to allow all members of the
community access to HST and JWST simulations, we present here an open source python package (also available as an online tool\footnote{\url{http://pandexo.science.psu.edu:1111}})
for creating observation simulations of all observatory-supported time-series
spectroscopy modes, called \texttt{PandExo}.
This noise simulator uses portions of \emph{Space Telescope
Science Institute}'s (STScI) Exposure Time Calculator, named \texttt{Pandeia}. We briefly
describe \texttt{Pandeia} in $\S$2 and how it is utilized within \texttt{PandExo} in
$\S$3. In $\S$4, we baseline \texttt{PandExo}'s performance against the JWST instrument team's simulators in order to show that they are in agreement. In $\S$5 we describe the methodology for simulating HST observations. We end with concluding remarks in $\S$6. 

\section{\texttt{Pandeia}: Simulating Noise Sources}
The source code for STScI's exposure time calculator, named \texttt{Pandeia}, was recently released to the 
observing community \footnote{\url{http://jwst.etc.stsci.edu}}.
Although \texttt{Pandeia} supports all officially-supported observing
modes, we limit our discussion to the modes that will be useful
for exoplanet transit spectroscopy.

\texttt{Pandeia} is a \emph{hybrid} instrument simulator. It simulates observations 
using a three-dimensional, pixel-based approach but its ultimate goal is to 
provide the user with accurate predictions of SNRs for specific observing scenarios. 
Therefore, it does not fully simulate the entire field of view of the 
instrument, and it does not include optical field distortion, intra-pixel 
response variations,  other detector systematic noise, 
or the effects of spacecraft jitter and drift.  \texttt{Pandeia} does include
accurate and up-to-date estimates for background noise, point-spread-functions (PSFs),
instrument throughputs and optical paths, saturation levels, ramp noise, 
correlated read noise, flat field errors and data extraction for all the JWST instruments. We briefly describe \texttt{Pandeia} below, but a full description can be found in \citet{pon16}. 

For each calculation, a three dimensional cube is created with spatial
and spectral dimensions. Astronomical scenes are modeled by specifying a
spectral energy distribution along these two dimensions. In the case of 
transit spectroscopy, this is always a stellar spectrum or star$+$planet
spectrum placed at the center of the optical axis and normalized at a 
specific reference wavelength (see \S3). 
\begin{figure}[ht]
 \includegraphics[angle=270,origin=c,width=\linewidth]{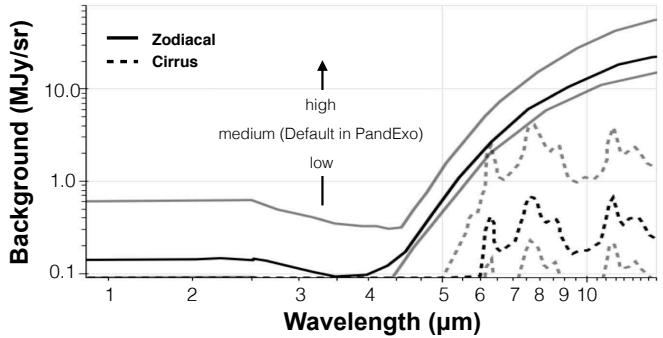}
\caption{Varying levels (low, medium, high) of pre-computed background flux (mega-Jy) used within \texttt{Pandeia} for cirrus (dashed) and zodiacal (solid) background contamination. Black curve shows the level used (medium) for all noise simulations in this analysis. 
\label{fig1}}
\end{figure}
After the scene is created, \texttt{Pandeia} uses pre-calculated low, medium 
and high background cases adopted from \citet{gla15}. For the calculations in this analysis we employ the ``medium'' background
case shown in Figure \ref{fig1}. 

\begin{figure}[ht]
 \includegraphics[angle=270,origin=c,width=\linewidth]{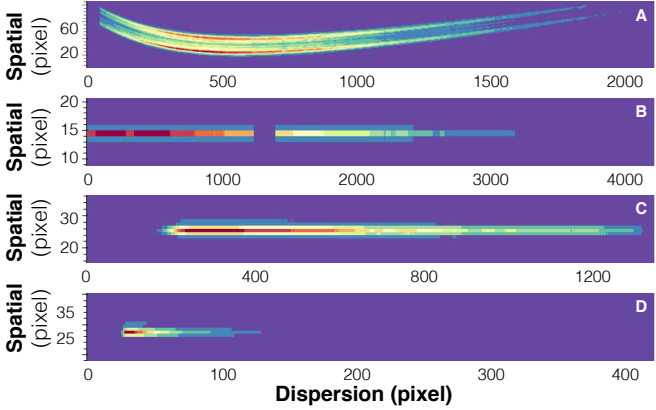}
\caption{Subsets of the two dimensional \texttt{Pandeia} detector simulation of a T=4000~K, Fe/H=0.0 and, log$g$=4.0 stellar SED normalized to a J=10, with 100 seconds of observing time. Color coding shows the electron rate in each pixel. Panel A is simulation of NIRISS SOSS (Order 2 not shown), panel B is a simulation of NIRSpec G395H, panel C is a simulation of NIRCam F444W, and panel D is a simulation of MIRI LRS. \label{fig2}}
\end{figure}

After the background is added, \texttt{Pandeia} convolves each plane in the 
three-dimensional astronomical scene with the unique, two dimensional PSF for 
the instrument mode being simulated. All PSFs are calculated using \emph{WebbPSF},
which is described in \citet{per11}. For spectroscopy modes (except NIRISS), \texttt{Pandeia} assumes 
that the PSF profile is independent of spatial location. The inclusion of the PSFs can be seen in \texttt{Pandeia}'s 2-dimensional simulations of the detector (Figure~\ref{fig2}). 
All JWST exoplanet time series spectroscopy modes
will acquire sampled-up-the-ramp data at a constant cadence of one \emph{frame} \citep{rau07}.
A frame is a unit of data that results from sequentially clocking and
digitizing all pixels in the rectangular area of the detector. The time it takes to
read out one frame ($t_f$) depends on the observation mode or, more specifically, on the
subarray size. In JWST terminology, a \emph{group} is a number ($n$) of consecutively
read frames with no intervening resets. For all exoplanet time series
modes, there is one frame per group. An \emph{integration}
is composed of a reset of the detector followed by a series of non-destructively sampled
groups (\emph{n} = \# groups per integration). The time it takes to reset the
detector in between integrations is equivalent to the frame time, $t_f$. 

The measured signal can be calculated in two ways. The first, referred to as MULTIACCUM, is the standard procedure within \texttt{Pandeia}. It computes the final signal by fitting each point up the ramp. The second, referred to as Last-Minus-First (LMF) is the standard procedure within \texttt{PandExo}. In this procedure, the final signal
within an integration is equal to the final readout value minus the 
first readout value. We describe MULTIACCUM below and describe LMF in \S3, where we also discuss differences between the two methods. 

In the MULTIACCUM procedure, correlations between the number
of groups and the number of averaged frames per group are considered when computing the individual noise on a single pixel. Generally, this data is modeled using a standard two-parameter least-squares fitting procedure. \citet{rau07} generalized this least-squares approach for fitting nondestructive reads for the JWST readout mode, accounting for the correlated noise in the integrating charge. \texttt{Pandeia} calculates the total noise via this formula \citep{rau07}: 
\begin{equation}
\begin{split}
        \sigma^2_\textrm{tot}  =  \frac{12(n-1)}{mn(n+1)}\sigma_\textrm{read}^2  &+\frac{6(n^2 +1)}{5n(n+1)}(n-1)t_gf   \\ & -  \frac{2(m^2-1)(n-1)}{mn(n+1)}t_ff 
\end{split}
\end{equation}
where $m$ is the number of frames per group (for transit time series $m=1$), 
$n$ is the number of groups, $\sigma_\textrm{read}$ is the read noise per frame, 
$t_g$ is the time per group, $f$ is the electron rate calculated from the 
astronomical scene cube ($e^{-1}$s$^{-1}$pixel$^{-1}$) and $t_f$ is the time per frame. 

The read noise, $\sigma_\textrm{read}$ (e$^-$ rms), is calculated by considering the 
effects of correlated noise. It is well-known that both the near-infrared 
H2RG detectors and the mid-infrared detectors are affected by correlated noise. 
Therefore, regardless of the amount of incident light on the detector, the 
read noise in one pixel will depend on the read noise in other pixels. This 
effect becomes even larger in the fast-read direction and ignoring it would 
lead to an underestimation of the noise. The greatest consequence of adding 
correlated noise is that the error propagation must be handled with a 
covariance matrix and the noise in each pixel cannot simply be added in 
quadrature sum. 

\section{\texttt{P\MakeLowercase{and}E\MakeLowercase{xo}}: Simulating JWST Observations}
Our JWST transit simulator tool, called \texttt{PandExo}, is built around the core capabilities of \texttt{Pandeia}'s throughput calculations. \texttt{Pandeia} is packaged as a
Python package that is called by \texttt{PandExo}, and therefore any updates to \texttt{Pandeia}, by STScI, will automatically (assuming user keeps python packages updated) be incorporated into \texttt{PandExo}.
In addition to the observatory inputs required for \texttt{Pandeia}, \texttt{PandExo} requires:
\begin{itemize}
    \item A stellar SED model ($F_{*,\lambda}$, taken from Phoenix Stellar Atlas \citep{huss13})
    \item Apparent magnitude 
    \item Planet spectrum (primary or secondary)
    \item Transit duration($T_{14}$)
    \item Fraction of time spent observing in-transit versus out-of-transit 
    \item Number of transits 
    \item Exposure level considered to be the saturation (\% full well)
    \item User-defined noise floor
\end{itemize}

Using the star and planet models, an out-of-transit ($F_{*,\lambda}$) and in-transit
model ($F_{*,\lambda}(1-(R_{p,\lambda}/R_*)^2)$ for primary transit or $(F_{p,\lambda} +
F_*)$ for secondary transits) is calculated. \texttt{PandExo} does not create full light curve
models with an ingress and egress. Likewise, it does not include the effects of time-varying
stellar noise. Doing so would require frame-by-frame simulations and would be too
computationally demanding for a community tool. We leave an in-depth analysis of these
effects for a future paper and treat the transit as a box model.  

With the out-of-transit spectrum, \texttt{PandExo} calls \texttt{Pandeia} to create a 2D simulated 
image of the flux on the detector with \emph{n} = 2 (minimum number of groups required
for an observation). Then, \texttt{PandExo} calculates the 
maximum number of groups allowed in an integration before the pixel on 
the detector receiving the highest flux reaches the user-defined saturation limit.
Determining how many groups per integration is a crucial step within \texttt{PandExo} because 
it sets the observing efficiency, also known as the duty cycle, where: 
\begin{equation}
    \textrm{eff} = \frac{n-1}{n+1} 
\end{equation}
The above equation is exact for the near-IR detectors, but MIRI is more efficient. MIRI reads pixels in two rows and then resets the two rows before going on to the next two rows. This dramatically shortens the dead time between the last read and reset (the denominator is only a little more than $n$ and little less than $n+1$. Therefore, while this is formula is exact for the near-IR instruments, MIRI is somewhat more efficient at small $n$ and so in that regime, PandExo values for MIRI may be slightly conservative.

After the timing information is calculated, \texttt{PandExo} uses \texttt{Pandeia} to compute two simulated extracted spectral rates (e$^- s^{-1}$): one for the out-of-transit component and one for the in-transit component. As discussed in $\S$2, \texttt{Pandeia} returns (among other
products) a 1D extracted flux rate ($F_{in,\lambda}$,$F_{out,\lambda}$ in e$^-$ s$^{-1}$). 

If there are $n_{i,in}$ integrations taken in, and $n_{i,out}$ integrations taken out of transit, the pure shot noise can be easily calculated from those fluxes via: 
\begin{table*}[ht]
\begin{center}
\caption{Instrument modes}
\begin{tabular}{lllll}
\tableline \tableline

Instrument      & Filter    & Wavelength Range  & Resolving Power   & RN   \\
                &           & ($\mu$m)          &                   & e$^-$/frame \\
\hline
NIRISS SOSS     & --        & 0.6--2.8          &    700            & 11.55\\
NIRSpec Prism   & Clear     & 0.7--5            &    100            & 16.8\\
NIRSpec G140M/H   & F070LP    &  0.7--1.27         &   1000/2700     & 16.8\\
NIRSpec G140M/H   & F100LP    &  0.97--1.89         &   1000/2700     & 16.8\\
NIRSpec G235M/H   & F170LP    &  1.70--3.0        &   1000/2700     & 16.8\\
NIRSpec G395M/H   & F290LP    &  2.9--5           &   1000/2700     & 16.8\\
NIRCam Grism    & F322W2    &  2.5--4.0         &   1500            & 10.96\\
NIRCam Grism    & F444W     &  3.9--5.0         &   1650            & 10.96\\
MIRI LRS        & --        &  5.0--14          &    100            & 32.6\\
WFC3 G102       & --        & 0.84--1.13        &    210            & 20.0 \\
WFC3 G141       & --        & 1.12--1.65        &    130            & 20.0 \\
\tableline
\label{instruments}
\end{tabular}
\end{center}
\end{table*}
\begin{equation}
    \sigma_{shot}^2 = F_{in,\lambda} t_g (n-1) n_{i,in} + F_{out,\lambda} t_g (n-1) n_{i,out}
\end{equation}
One important correction that is made to $F_{in/out,\lambda}$ from \texttt{Pandeia}, is the contribution from quantum yield. Quantum yield is the number of charge carriers generated per interacting photon \citep{jan11}. It ultimately has the effect of increasing the electron rate and by default, the saturation rate, of the detectors by a factor of $\sim$1.8 at 0.5~$\micron$, dropping to a factor of $\sim$1.0 at 1.9~$\micron$ (for the nir-IR detectors) \citep{pon16}. In \texttt{Pandeia}, this is added to the extracted flux product and corrected for in the noise product. In \texttt{PandExo}, we divide \texttt{Pandeia}'s extracted flux by the quantum yield before computing $\sigma_{shot}^2$. 

Then, to compute the total noise, we must add in the contributions from the background and the read noise. The background signal, $F_{bkg}$, is directly computed from \texttt{Pandeia}. The contribution from the read noise is:
\begin{equation}
    \sigma_{read}^2 = 2\textrm{RN}^2 n_{pix} (n_{i,in} + n_{i,out})
\end{equation}
where RN is the total contribution of read noise in electrons (see Table 1), $n_{pix}$ is the number of extracted pixels. The factor of two comes from the fact that \texttt{Pandeia}'s RN values (Table 1) are given in units of e$^-$/frame. Since we are subtracting the last frame from the first, we must account for both frames.  The total noise, calculated for the in-transit and out-of-transit data separately, is then 
\begin{equation}
    \sigma_{tot}^2 = \sigma_{shot}^2 + \sigma_{bkg}^2 + \sigma_{read}^2.
\end{equation}
This traditional formulation does not assume any correlations between the number of groups. The total number of electrons collected and the associated noise is simply computed by subtracting the first group from the last group, LMF. The last group can always be used because \texttt{PandExo} does not model non-linearity. Once JWST's non-linearity is more accurately known \texttt{PandExo} will be updated accordingly. 

As discussed above, in \texttt{Pandeia}'s MULTIACCUM formulation (Eqn. 1) all the groups in the data are used to fit a slope. The first group has $t_g*F_{out}$ electrons, the second as $2t_g*F_{out}$ electrons, etc. And although each of these groups has a separate photon noise component, the noise is correlated between all of them. Therefore, the MULTIACCUM method will only be equivalent to the LMF method in the case where the flux rate, $F$, is much larger than the expected read noise and $n=2$. In this limit, Eqn 1 simplifies to: 
\begin{equation}
    \sigma_{tot,MULTI}^2 \approx t_g(F_{in}n_{i,in}  + F_{out}n_{i,out} )
\end{equation}
The number of groups will be $2$ in the cases where the magnitude of the target is very close to the saturation limit of the instrument mode, which will be the case for a small number of exoplanet targets. In these cases, the MULTIACCUM method and the LMF method will yield similar results, barring small correlated noise contributions from the MULTIACCUM method. However, in cases where the magnitude of the target is at least $\sim1$ magnitude greater than the saturation limit of the instrument mode, $n$ will be much larger than 2 and the flux rate will still be much larger than the expected read noise. In this limit, the MULTIACCUM formulation simplifies to:
\begin{equation}
    \sigma_{tot,MULTI}^2 \approx \frac{6}{5}t_gn (F_{in}n_{i,in}  + F_{out}n_{i,out}).
\end{equation}
In this limit, the factor of $\frac{6}{5}$ comes from the second expression in Eqn. 1. When $n=2$, $\frac{6(n^2+1)}{5n(n+1)} = 1$, but the $\frac{6}{5}$ factor remains when $n>2$. Therefore in the $n>2$ regime, the uncertainty calculated using the MULTIACCUM method will be a factor $\sim \frac{6}{5}$ greater than the LMF method (Eqn. 3). 

To reconcile these two different noise formulations, \texttt{PandExo} has the capability to derive the noise using either method by simply changing a key word in the input file. However, the default noise calculation is the simplified LMF method. 

The final simulated transmission ($-$) and emission ($+$) spectra combines these and 
adds a random noise component via the equation:
\begin{equation}
    z_\lambda = 1 \pm \frac{N_{in,\lambda}}{N_{out,\lambda}} + \sqrt{\sigma_{prop}^2} \times N(0,1)
\end{equation}
where $N_{out/in,\lambda}$ is the total number of photons collected out-of- and in-transit and $N(0, 1)$ is a standard normal distribution. The 1$\sigma$ propagated error on the final spectrum, $\sigma_{prop}$ is: 
\begin{equation}
    \sigma_{prop}^2 = \sigma_{in}^2 \left( \frac{n_{i,out}}{n_{i,in}N_{out,\lambda}} \right)^2 + \sigma_{out}^2 \left( \frac{n_{i,out}N_{in,\lambda}}{n_{i,in}N_{out,\lambda}^2} \right)^2
\end{equation}
\begin{figure}[ht]
 \includegraphics[angle=270,origin=c,width=\linewidth]{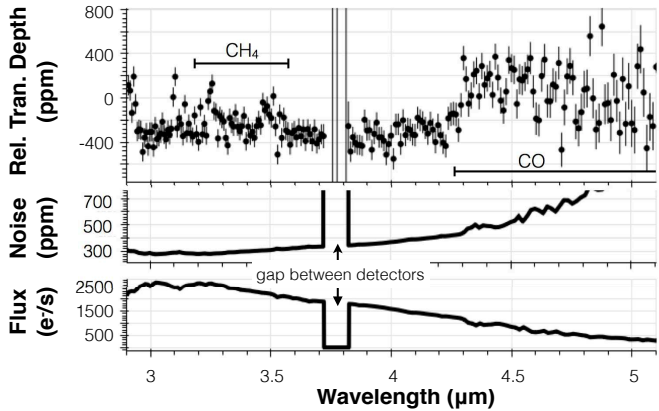}
\caption{Three of the most popular \texttt{PandExo} output products. The top panel is the raw planet transmissions spectrum with associated errors. The middle panel is the raw noise and the bottom panel is the out of transit flux rate. Each simulation is for a NIRSpec G395H observation of a T=4000~K, Fe/H=0.0 and log$g$=4.0 stellar target normalized to a J=10. The single transit observation consists of a 2.7 hour in-transit observation along with a 2.7 out-of-transit baseline observation (chosen for simplicity). In the top panel, the observation is binned to R=200. In the middle and bottom panel, resolving power is left at native resolving power (per pixel), with ten pixels summed in the spatial direction. The gaps seen in all three panels are the result of a gap between the detectors from 3.8172-3.9163~$\micron$. \label{fig3}}
\end{figure}
\citet{gre16} argue that a systematic noise floor might inhibit JWST observations to get
below 20 ppm, 30 ppm, and 50 ppm, for NIRISS SOSS, NIRCam grism, and MIRI LRS,
respectively. This argument was based off of a comparison of the lowest noise achieved with
an HST WFC3 G141 observations \citep{kre14} for the NIR instruments and the lowest noise
achieved with a Spitzer Si:As observation \citep{knu09} for MIRI.
However, as with the actual effective saturation limit, the noise floor ($\sigma_{f,\lambda}$) 
will not be known until after commissioning and Early Release Science \citep{ste16}.
Therefore we do not adopt these same noise floors and leave it up to the observer to input
their own. In contrast to \citet{gre16}, noise floors are not added to
$\sigma_{prop,\lambda}$ in quadrature. Instead, \texttt{PandExo} sets
$\sigma_{prop,\lambda}(\sigma_{prop,\lambda}<\sigma_{f,\lambda})=\sigma_{f,\lambda}$. 
This is done solely to increase the transparency of the calculation. The major final \texttt{PandExo} products are shown in Figure~\ref{fig3}. 

\section{Benchmarking \texttt{P\MakeLowercase{and}E\MakeLowercase{xo}} Performance}
\begin{figure}[ht]
 \includegraphics[angle=270,origin=c,width=\linewidth]{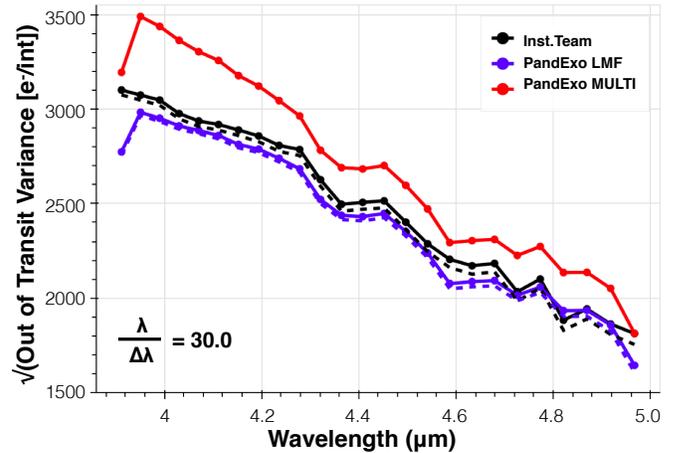}
\caption{Benchmarking results for NIRCam, which show the differences between the two \texttt{PandExo} noise formulations and the instrument team's simulations. The specific observing mode depicted is NIRCam F444W, which was run with one integration both in- and out-of-transit. In solid black is the instrument team's noise simulation, which includes all pertinent sources of noise. In solid blue and red is \texttt{PandExo}'s LMF and MULTIACCUM noise formulation, respectively (see discussion in \S3). In dashed blue and black is the instrument team's and \texttt{PandExo}'s calculation for pure shot noise, respectively.\label{fig4}}
\end{figure}

\begin{figure}[ht]
 \includegraphics[angle=270,origin=c,width=\linewidth]{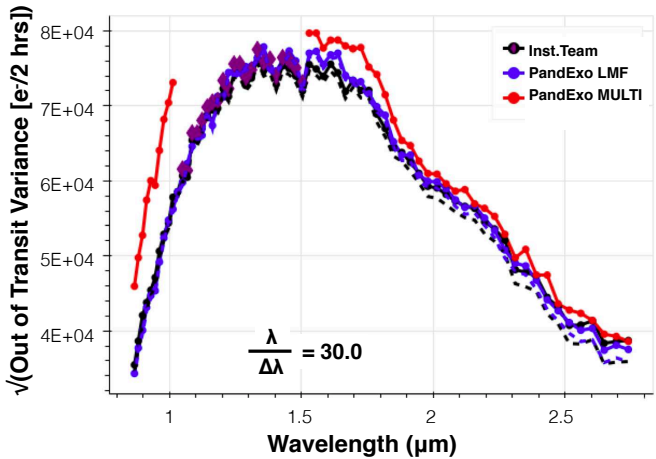}
\caption{Benchmarking results for NIRISS, which show the differences between the two \texttt{PandExo} noise formulations and the instrument team's simulations. The specific observing mode depicted is NIRISS SOSS, which was run for a 4 hour integration (2 hours in- and 2 hours out-of-transit) with 2 groups per integration. In solid black is the instrument team's noise simulation, which includes all pertinent sources of noise. In solid blue and red is \texttt{PandExo}'s LMF and MULTIACCUM noise formulation, respectively (see discussion in \S3). In dashed blue and black is the instrument team's and \texttt{PandExo}'s calculation for pure shot noise, respectively. Missing points in red depict pixels that have been saturated in \texttt{PandExo}. Likewise, purple diamonds depict pixels that have been saturated in the instrument team's model. \label{fig5}}
\end{figure}

In the absence of JWST observations, we test the accuracy of \texttt{PandExo} against each instrument team's independently written noise simulators. Each of the instrument teams used the LMF noise formulation. For completeness we show the LMF (always in blue) and the MULTIACCUM noise derivations (always in red) as well as the pure shot noise (always dashed lines). The following calculations are also all done using a stellar SED from the Phoenix Stellar Database \citep{huss13} with T=4000~K, Fe/H=0.0 and, log$g$=4.0 normalized to a J=8, a model of WASP-12b in transmission from \citet{mad14}, and $n_{i,in}=n_{i,out}$ (chosen for simplicity). The results of the comparisons are shown in Figure \ref{fig4}, \ref{fig5}, \ref{fig7}, and \ref{fig8}, and discussed in the following sections. 
\subsection{NIRCam}
To benchmark NIRCam, we used the NIRCam F444W grism 
mode in conjunction with the SUBGRISM64 subarray ($t_g$=0.34~secs). The results are shown in Figure~\ref{fig4}. For a target with J=8, we selected $n=55$ groups to optimize the duty cycle without saturating the detectors (eff=0.96) and ran the simulation for a single integration in transit and a single integration out of transit. Because of the high number of groups, \texttt{PandExo} MULTIACCUM, as expected, is a factor of $\sim \frac{6}{5}$ higher than \texttt{PandExo} LMF (see discussion in $\S3$) and \texttt{PandExo} LMF, as expected, matches within 10\% with the instrument team's results. 

NIRCam is also slitless. While PandExo does not directly incorporate position angles to prevent overlapping spectra, it is important to consider this when planning observations. 
\subsection{NIRISS}
To benchmark NIRISS, we used the NIRISS SOSS mode in conjunction with the SUBSTRIP256 subarray ($t_g$=5.491~secs). The results are shown in Figure~\ref{fig5}. For a target with J=8, only the minimum number of groups, $n=2$, is possible. And even so, this results in a partial saturation of pixels at the peak of the stellar SED. NIRISS simulations were computed with a 2 hour observation in transit, and 2 hour baseline observation out of transit. 

Because $n=2$, the MULTIACCUM (red) approximately follows the \texttt{PandExo} LMF (blue). The omitted points in the red curve, and the purple diamonds represent saturated pixels in \texttt{PandExo} and the instrument team's simulator, respectively. Both teams are saturating identical pixels. 

\begin{figure}[ht]
 \includegraphics[angle=270,origin=c,width=\linewidth]{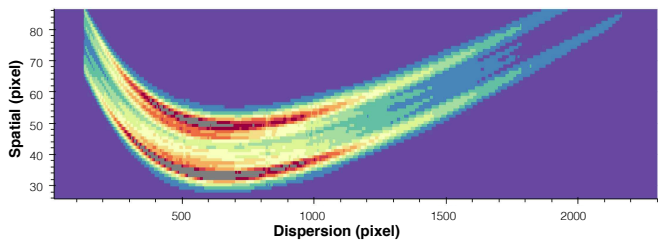}
\caption{2D detector simulation for the NIRISS SOSS observation shown in Figure~\ref{fig5}. Only the first order is depicted to enable a clear view of the saturated pixels (colored in grey). Color indicates electron rate in e$^-$ s$^{-1}$. The wavelength channels with saturated pixels are flagged by \texttt{PandExo} but usable data may still be extractable from non-saturated regions.\label{fig6}}
\end{figure}
\texttt{PandExo} produces 2-dimensional simulations of detector images and of the saturation profiles. Figure~\ref{fig6} shows the exact pixels that saturated, colored in gray. Because of NIRISS' widely sampled PSF (23 pixels), it is likely still possible to extract a spectrum by excluding saturated pixels (a decision the observer must make). Ultimately though, if an observation only contains 2 groups, \texttt{PandExo} marks every wavelength bin which contains at least one saturated pixel as completely saturated, regardless of whether or not it may be possible to extract unsaturated data from that bin. \texttt{PandExo} will then produce the following warning statement: ``There are \emph{[\# OF PIXELS]} saturated pixels at the end of the first group. These pixels cannot be recovered.''  The NIRISS team also alerts users by flagging each pixel considering saturated. By adding in these obvious warnings, users will know they are in a region of parameter space where they will, to some degree, encounter saturated pixels. 

An important limitation with NIRISS is contamination by field stars because it is slitless. It is crucial to run the instrument team's contamination tool to select observing position angles and dates that minimize spectral trace contamination. It complements the instrument team's 1D simulator \footnote{\url{http://jwst.astro.umontreal.ca/?page\_id=401}} used for comparison with \texttt{PandExo}.
\begin{figure}[ht]
 \includegraphics[angle=270,origin=c,width=\linewidth]{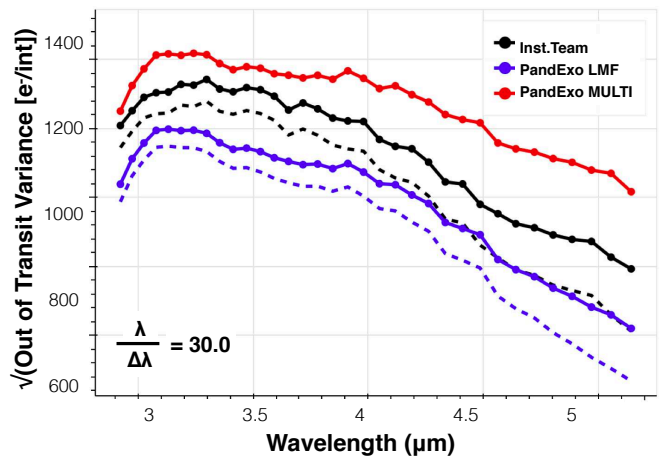}
\caption{Benchmarking results for NIRSpec, which show the differences between the two \texttt{PandExo} noise formulations and the instrument team's simulations. The specific observing mode depicted is NIRSpec G395H with the f090lp filter, which was run for a single integration in- and out-of-transit. In solid black is the instrument team's noise simulation, which includes all pertinent sources of noise. In solid blue and red is \texttt{PandExo}'s LMF and MULTIACCUM noise formulation, respectively (see discussion in \S3). In dashed blue and black is the instrument team's and \texttt{PandExo}'s calculation for pure shot noise, respectively. \label{fig7}}
\end{figure}
\subsection{NIRSpec}
To benchmark NIRSpec, the G395M/F290LP grism/filter was used with the the 32x2048 subarray ($t_g=0.90156$~secs). The results are shown in Figure~\ref{fig7}. For the benchmarking, we simulated an observation with $n=2$ for a single integration out-of-transit and a single integration in-transit. The instrument team's simulator, described in \citet{nie16}, can either implement the LMF noise procedure or the "Last-Minus-Zero" (LMZ) procedure. In this strategy, the observer implements a reset-read-reset scheme with $n=1$. Currently in \texttt{PandExo} the number of groups must be $n\ge2$. This requirement is a result of \texttt{Pandeia}'s requirements and will be lifted as soon as \texttt{Pandeia} is updated. Here, we only consider the instrument team's LMF noise formula. As expected, these match within 10\%.
\begin{figure}[ht]
 \includegraphics[angle=270,origin=c,width=\linewidth]{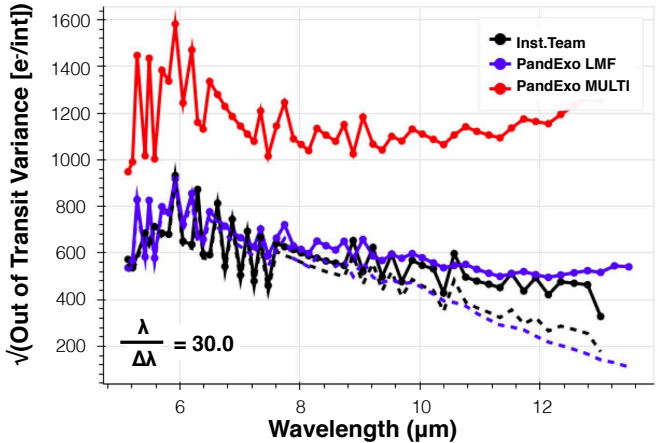}
\caption{Benchmarking results for MIRI, which show the differences between the two \texttt{PandExo} noise formulations and the instrument team's simulations. The specific observing mode depicted is MIRI LRS slitless mode, which was run for a single integration in- and out-of-transit. In solid black is the instrument team's noise simulation, which includes all pertinent sources of noise. In solid blue and red is \texttt{PandExo}'s LMF and MULTIACCUM noise formulation, respectively (see discussion in \S3). In dashed blue and black is the instrument team's and \texttt{PandExo}'s calculation for pure shot noise, respectively. \label{fig8}}
\end{figure}
\subsection{MIRI}
For MIRI, the LRS slitless mode was used ($t_g$=0.159~secs). Figure~\ref{fig8} shows the results. For a J=8 target, we selected $n=10$ groups to optimize the duty cycle without saturation (eff=0.81) and ran the simulation for a single integration in transit and a single integration out of transit. Similar to NIRCam, the MULTIACCUM \texttt{PandExo} results are offset by $\sim \frac{6}{5}$ because $n>2$. 

The \texttt{PandExo} LMF formulation results are in good agreement with the instrument team's simulations. It should be pointed out that the jagged behavior of the noise curve is solely a result of binning ($\lambda/\Delta \lambda = 30$ creates variable pixels per bin) and not an instrument systematic. Also, in both teams simulations MIRI is slightly dominated by read noise and background at long wavelengths ($\lambda>10\micron$) for a target with J=8. This adds another high degree of certainty to the correctness of both calculations. MIRI is also technically slitless, but the SLITLESSPRISM subarray is small enough so that overlapping spectra are not a major issue.
\begin{figure*}[ht]
\centering
 \includegraphics[angle=270,origin=c,width=6in]{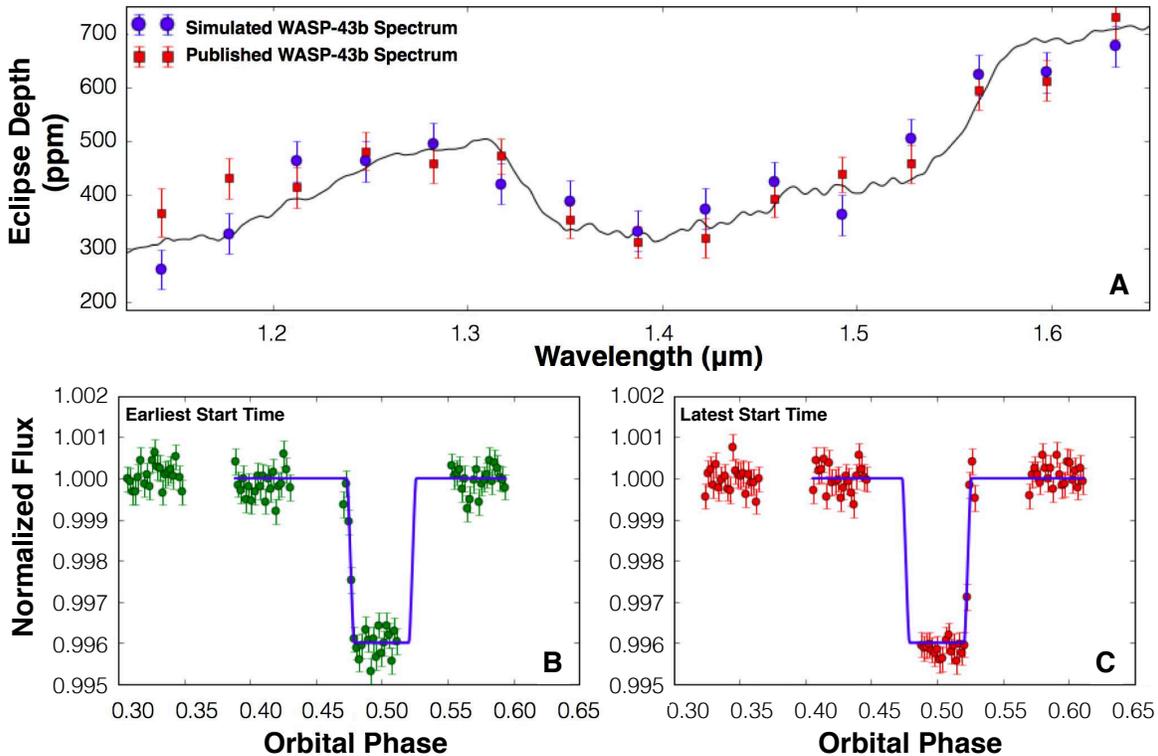}
\caption{Simulated observations of WASP-43b in emission using HST WFC3 G141. For the utilized instrument configuration, the simulated uncertainty is 37.6 ppm and the mean published uncertainty is $36.5 {\pm} 3.5$~ppm \citep{Stevenson2014c}. For the same example system, panels B \& C display simulated band-integrated light curves with the earliest and latest possible observation start times, respectively, that correspond to the computed minimum and maximum phase values of 0.3071 and 0.3241.\label{fig9}}
\end{figure*}
\section{\texttt{PandExo}: Simulating HST Observations}
In addition to simulating JWST observations, \texttt{PandExo} can simulate realistic uncertainties for HST/WFC3 transmission and emission spectra, optimize instrument setups, and generate scheduling requirements.  Accurate spectrophotometric uncertainties are necessary to correctly determine the number of transit/eclipse visits required to obtain a meaningful constraint.

The HST/WFC3 implementation of \texttt{PandExo} predicts spectrophotometric uncertainties for any specified system by first scaling measured flux, variance, and exposure time values from previously-observed systems published in \citet{Kreidberg2014a} and \citet{kre14}, then computing the expected rms per spectrophotometric channel per exposure, and finally estimating the transit/eclipse depth error based on the anticipated number of individual valid in- and out-of-transit exposures.  The uncertainty estimates depend on the orbital properties of the system, instrument configuration, and observation duration.  The code assumes Gaussian-distributed white noise and uniform uncertainties over the G102 and G141 grisms, both of which are consistent with published results \citep[e.g.][]{Kreidberg2014a, kre14, Stevenson2014c}.  \texttt{PandExo} also recommends an observing strategy (best NSAMP and SAMP-SEQ values) optimized to achieve the highest duty cycle (lowest photon-noise rms) and computes an observation start range in units of orbital phase.  These instrument and scheduling requirements are important factors to consider when planning proposals and observations in the Astronomer's Proposal Tools (APT) and can be tedious to compute/optimize manually for a large number of targets.

As inputs, \texttt{PandExo} requires the stellar H-band magnitude, full transit/eclipse duration, number of transits/eclipses, number of spectrophotometric channels, disperser type (G102 or G141), scan direction (forward or round trip), subarray size (GRISM256 or GRISM512), and schedulability (30\% for small/medium programs or 100\% for large programs).  Optional inputs that may be optimized include the number of HST orbits per visit and WFC3's instrument parameters (NSAMP and SAMP-SEQ).  Additional inputs for the scheduling requirement include the orbital parameters (transit/eclipse depth, inclination, separation, eccentricity, longitude of periastron, and period) and the observation start window size (usually 20 -- 30 minutes).

Accepting a user-provided model transmission/emission spectrum, \texttt{PandExo} will simulate binned spectrophotometric data with realistic uncertainties and plot the results against the supplied model.  As an example, Figure \ref{fig9}A depicts a model emission spectrum of WASP-43b at secondary eclipse as well as simulated and published WFC3/G141 data.  For the utilized instrument configuration, the simulated uncertainty is 37.6 ppm and the mean published uncertainty is $36.5 {\pm} 3.5$~ppm \citep{Stevenson2014c}.  These values are consistent at 0.3$\sigma$.

For the same example system, Figure \ref{fig9}B \& C display simulated WFC3 light curves with the earliest and latest possible observation start times, respectively, that correspond to the computed minimum and maximum phase values of 0.3071 and 0.3241, respectively.  The actual observations would commence anywhere in between these two extremes.

Future work for this noise simulator includes adding functionality for the STIS G430 and G750 grisms, computing wavelength-dependent uncertainties, and exploring more sophisticated calculation methods beyond scaling values from previously-observed systems.

\section{Conclusion}
We introduced a new open-source Python package, called \texttt{PandExo}, which is used for modeling instrumental noise from each of the exoplanet transit time series exoplanet spectroscopy modes with JWST (NIRISS, NIRCam, NIRSpec and, MIRI LRS) and HST (WFC3). \texttt{PandExo} computes noise with two different noise formulations: 1) subtracting the last group from the first group (LMF method) and 2) independently fitting each group up the ramp (MULTIACCUM method) and accounting for correlated noise. The instrument teams' calculations are in good agreement with \texttt{PandExo}'s noise calculations employing the LMF method.

\texttt{PandExo} currently does not include any photometry modes. However, it is expected to continue evolving as we approach JWST launch date in 2018. 

\texttt{PandExo} is available for download on github\footnote{\url{https://github.com/natashabatalha/PandExo}} and there is an associated github pages with full documentation and tutorials \footnote{\url{https://natashabatalha.github.io/PandExo}}. The online interface is also currently available \footnote{\url{https://pandeexo.science.psu.edu:1111}}. 

\acknowledgments
\section{Acknowledgments}
First and foremost, we thank the STScI team responsible for writing and developing \texttt{Pandeia}. This was a huge undertaking which will undoubtedly improve the astrophysics community's understanding of JWST. We also thank Hannah R. Wakeford and Sarah Blumenthal for countless discussions regarding how to create the most useful and user-friendly tool. Lastly, we thank Natalie M. Batalha, Laura Kreidberg, Ian Crossfield, Kamen Todorov, Nicolas Crouzet and Zach Berta-Thompson for testing \texttt{PandExo}, which resulted in bug fixes and code/installation improvements. This material is based upon work supported by the National Science Foundation under Grant No. DGE1255832 to N.E.B. Any opinions, findings, and conclusions or recommendations expressed in this material are those of the author(s) and do not necessarily reflect the views of NASA or the National Science Foundation.

\end{document}